\documentclass[a4paper]{jpconf}
\pdfoutput=1
\usepackage{graphicx}
\usepackage{subfigure}
\begin{document}

\newcommand{\met}{\ensuremath{{\not\mathrel{E}}_T}}

\title{Single Top Quark Measurements at the Tevatron}

\author{Manfredi Ronzani (on behalf of the CDF and D0 collaborations)}

\address{Albert-Ludwigs-Universit\"{a}t Freiburg, Physikalisches Institut, Hermann-Herder-Stra\ss e 3a, D-79104 Freiburg, Germany}

\ead{manfredi.ronzani@cern.ch}
\begin{abstract}
This paper reports the most recent measurements of single top quark production performed by CDF and D0 collaborations in proton-antiproton collisions at Tevatron. Events are selected in the lepton+jets final state by CDF and D0 and in the missing transverse energy plus jets final state by CDF. The small single top signal in s-channel, t-channel and inclusive s+t channel is separated from the large background by using different multivariate techniques. We also present the most recent results on extraction of the CKM matrix element $|V_{tb}|$  from the single top quark cross section. 
\end{abstract}

\section{Introduction}
The Fermilab Tevatron Collider was in operation until September 2011 and in the RunII it provided 12 fb$^{-1}$ of $\rm p\bar{p}$ collisions at center of mass energy of $\sqrt{s}$ = 1.96 TeV. The D0 and CDF detectors recorded ~10 fb$^{-1}$ of proton-antiproton data per experiment. 

\begin{figure}[t]
\begin{center}
\includegraphics[width=23pc]{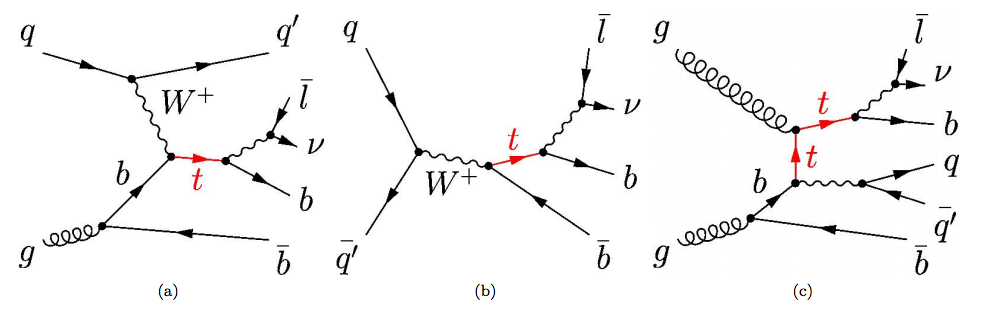}
\caption{\label{single_top_dyagrams}Single top quark production modes: (a) t-channel at NLO with initial state gluon splitting, (b) s-channel at leading order, and (c) Wt-channel at NLO with initial state gluon splitting.}
\end{center}
\end{figure}

The top quark was observed at the Tevatron by CDF \cite{Abe} and D0 \cite{Abachi} in 1995 in $t\bar t$ pairs produced via strong interaction that is the top quark primary production mode. The Standard Model (SM) predicts top quark to be produced also singly via electroweak interaction (single top) with three different production processes (Fig.\ref{single_top_dyagrams}): t-channel with the exchange of a virtual W boson \cite{t-channel}, s-channel with the W boson decaying into a top and an antibottom quarks \cite{s-channel} and the Wt-channel with the associate production of a W boson and a top quark \cite{Wt-channel}. While the first two production modes have a small but measurable cross section at Tevatron (around 2 pb and 1 pb, respectively) the Wt production mode has a cross section of 0.25 pb, that makes its contribution to the signal rate negligible. 

Single top quark was first observed in 2009 by CDF \cite{CDF2009} and D0 \cite{D02009} experiments with an inclusive search in s+t combined channels. The measurement in each single channel is done differently at the Tevatron and LHC, taking in account the differences in the production cross sections \cite{t-channel,s-channel,Wt-channel}. The t-channel is the dominant process at both the Tevatron and LHC and it has been observed by D0 in 2011 \cite{D0_t-channel} and then established by ATLAS \cite{ATLAS_t-channel} and CMS \cite{CMS_t-channel}. The Wt-channel is visible only at the LHC and it has been observed by CMS collaboration at 8 TeV \cite{CMS_Wt-channel}. The s-channel has a relatively small cross section at both Tevatron and LHC but at LHC the signal to background ratio is smaller, that gives an advantage for the observation to the Tevatron experiments. 

The single top quark cross section measurement is  proportional to $\rm |V_{tb}|^2$, where $\rm |V_{tb}|$ is element of the Cabibbo-Kobayashi-Maskawa (CKM) matrix. A direct measurement of $\rm |V_{tb}|$ is therefore possible, that is also a test of the unitarity of the CKM matrix and it can constrain extensions of the SM, for example with fourth quark generation \cite{Vtb_theory}.

\section{Selection of single top quark events and modeling}
The SM predicts the top quark to decay almost always into a real $W$ boson and a $b$ quark. The further decay of the $W$ boson selects the final state. Both CDF and D0 use lepton+jets final state selection requiring the presence of an high-$\rm p_{T}$ isolated lepton (electron or muon), large missing transverse energy (\met) and 2 or 3 jets, at least one of them identified as coming from a b quark (1 or 2 b-tags). At CDF, the \met+jets final state selection is also investigated.
In the \met+jets selection the products from the decay of the W boson are not reconstructed or they contain $\tau$ leptons which decay hadronically. This selection is orthogonal to the lepton+jets one and adds approximately 33\% of acceptance. 


Different background processes 
are considered: events with a W boson produced with one or more heavy flavor jets (W+HF) or with light-flavor jets that are mistagged as coming from a b quark (W+LF), events from diboson processes (WW,ZZ,WZ), events with a Z boson and jets, $t\bar t$ pair production and multijets (non-W) events where a W boson is falsely reconstructed. All the background processes (except multijets) and the single top signal events are modeled using various MC event generators. Single top signals are modeled for a top quark mass of $m_t = 172.5$ GeV/c$^2$ using {\rmfamily\scshape powheg} (CDF) \cite{Powheg} or {\rmfamily\scshape comphep} (D0 )\cite{CompHEP}. The multijet background is modeled using data. 

At the end of the cut-and-count analysis the signal over background ratio is still of the order of 1/20. Moreover, the number of expected signal events is much smaller than the uncertainty of the predicted background, that makes a precise measurement of the single top cross section not possible. Different multivariate analysis methods (MVA) and combination of different MVAs are used to extract the signal from the background.

\section{Single top production cross-section measurement}

\subsection{CDF: s+t  single top analyses}
\begin{figure}[ht]
\centering
\subfigure[]{
\includegraphics[width=11pc]{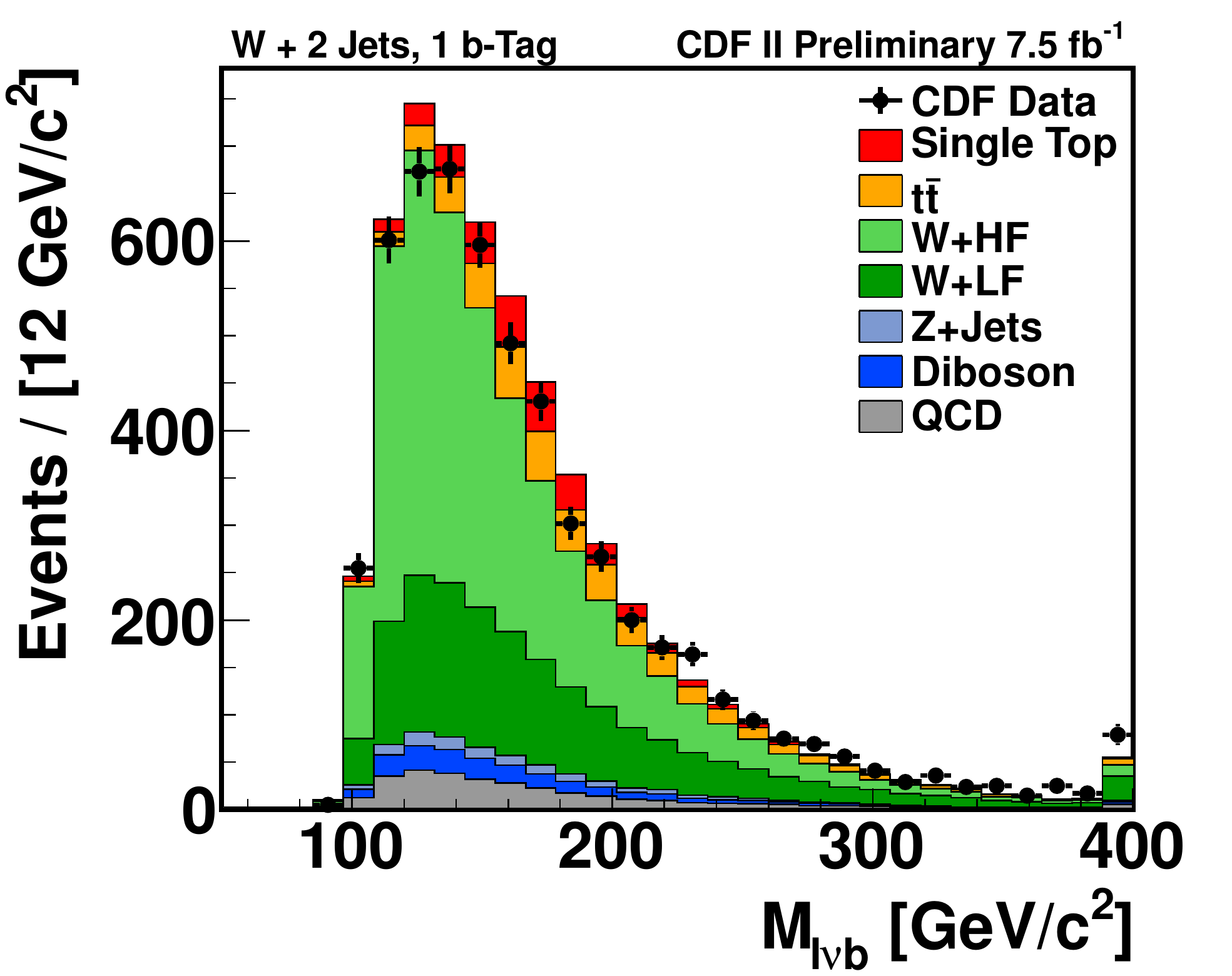}
\label{fig:2J1T_MlnubSum_Stacked_NormPred_PRD}}
\quad
\subfigure[]{
\includegraphics[width=11pc]{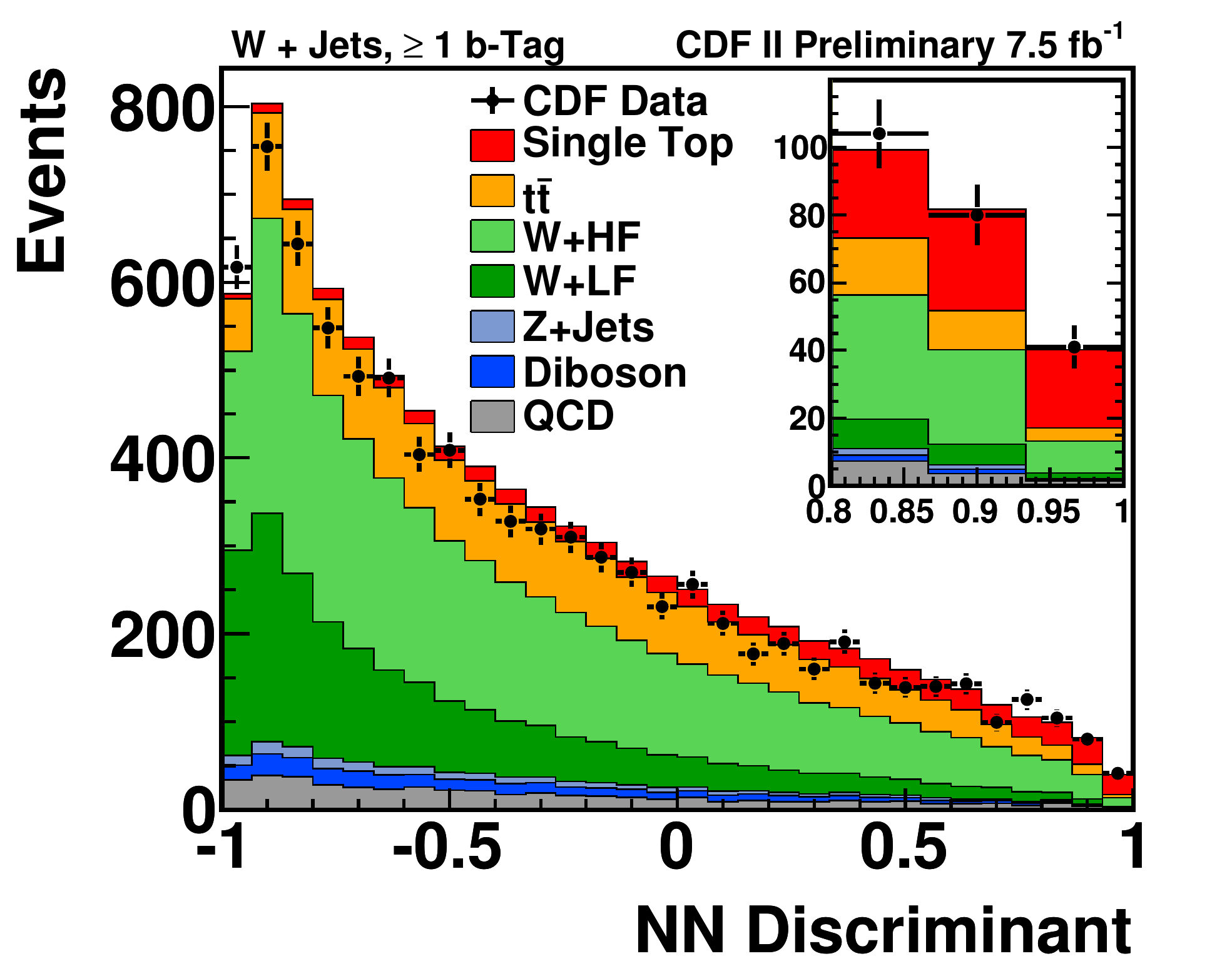}
\label{fig:all_NNoutSum_inlaidZoom_Stacked_NormPred_PRD}}
\caption{Reconstructed top quark mass for the 2J1T events (a) and the NN final discriminant (b) of the CDF l+jets analysis.}
\label{fig:figure}
\end{figure}

At CDF the s+t inclusive channel has been fully investigated in both l+jets and \met+jets final states. 
The l+jets analysis is performed in 7.5 fb$^{-1}$ of integrated luminosity. The analysis divides the selected sample into 4 independent signal regions depending from the jet multiplicity (2 or 3) and the number of b-tags (1 or 2). A Neural Network (NN) discriminant is built from a number of kinematic variables with good discriminating power (Fig.\ref{fig:2J1T_MlnubSum_Stacked_NormPred_PRD}) and it is trained with both s-channel (in 2J2T region) and t-channel (in the other regions) as signal. 
Fig.\ref{fig:all_NNoutSum_inlaidZoom_Stacked_NormPred_PRD} shows the data-prediction comparison of the final NN output in the combined signal regions. 
The total single top cross section value is measured to be 
$\rm \sigma_{s+t+Wt} = 3.04^{+0.57}_{-0.53}$ pb for a top quark mass of $\rm m_{t} = 172.5$ GeV/c$^2$. A 2-dimensional fit is performed in order to get the cross section for s-channel and t+Wt channel separately (Fig.\ref{fig:2D_sVst})\cite{st_CDF_lepjet}. The t and Wt channels are combined as they share the same final state topology. 

The \met+jets analysis is performed in 9.5 fb$^{-1}$ of CDF integrated luminosity. The analysis divides the selected sample in 6 subsamples depending on the number of jet multiplicity (2 or 3) and the tagging classification (tightly tag T or loosely tag L), which is based on a multivariate algorithm initially developed for the Higgs analysis. The background is composed mainly by multijets events and it is discriminated using several dedicated NNs. A signal NN$^{\rm sig}_{s+t}$ is used to separate both s- ant t-channel signal processes from the background. 
The measured inclusive cross section is $\sigma_{s+t} = 3.53^{+1.25}_{-1.16}$ pb \cite{st_CDF_metjet}.
 
The results of the two CDF s+t analyses (lepton+jets and \met+jets) are combined. A t-channel single top quark production cross section, considering the s-channel as a background constrained to the theoretical prediction, is measured to be $\sigma_t = 1.65^{+0.38}_{-0.36}$ pb. The combined $\sigma_{s+t}$ measurement results in a single top quark production cross section of $\sigma_{s+t} = 3.02^{+0.49}_{-0.48}$ pb, consistent with the SM prediction \cite{st_CDF_metjet}. 

\subsection{D0 s+t single top analysis}
\begin{figure}[ht]
\centering
\subfigure[]{
\includegraphics[width=11pc]{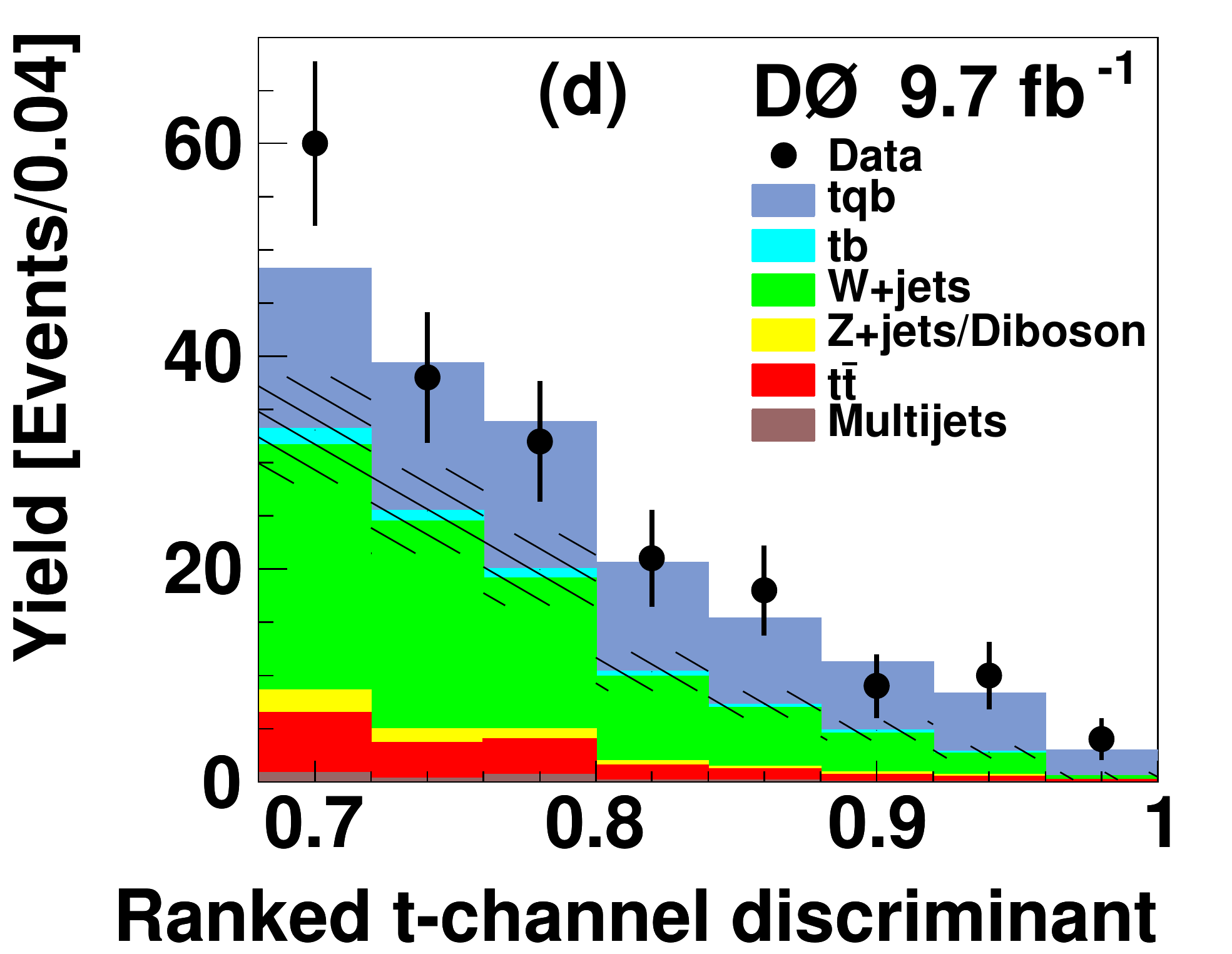}
\label{fig:T13AF3d}}
\quad
\subfigure[]{
\includegraphics[width=11pc]{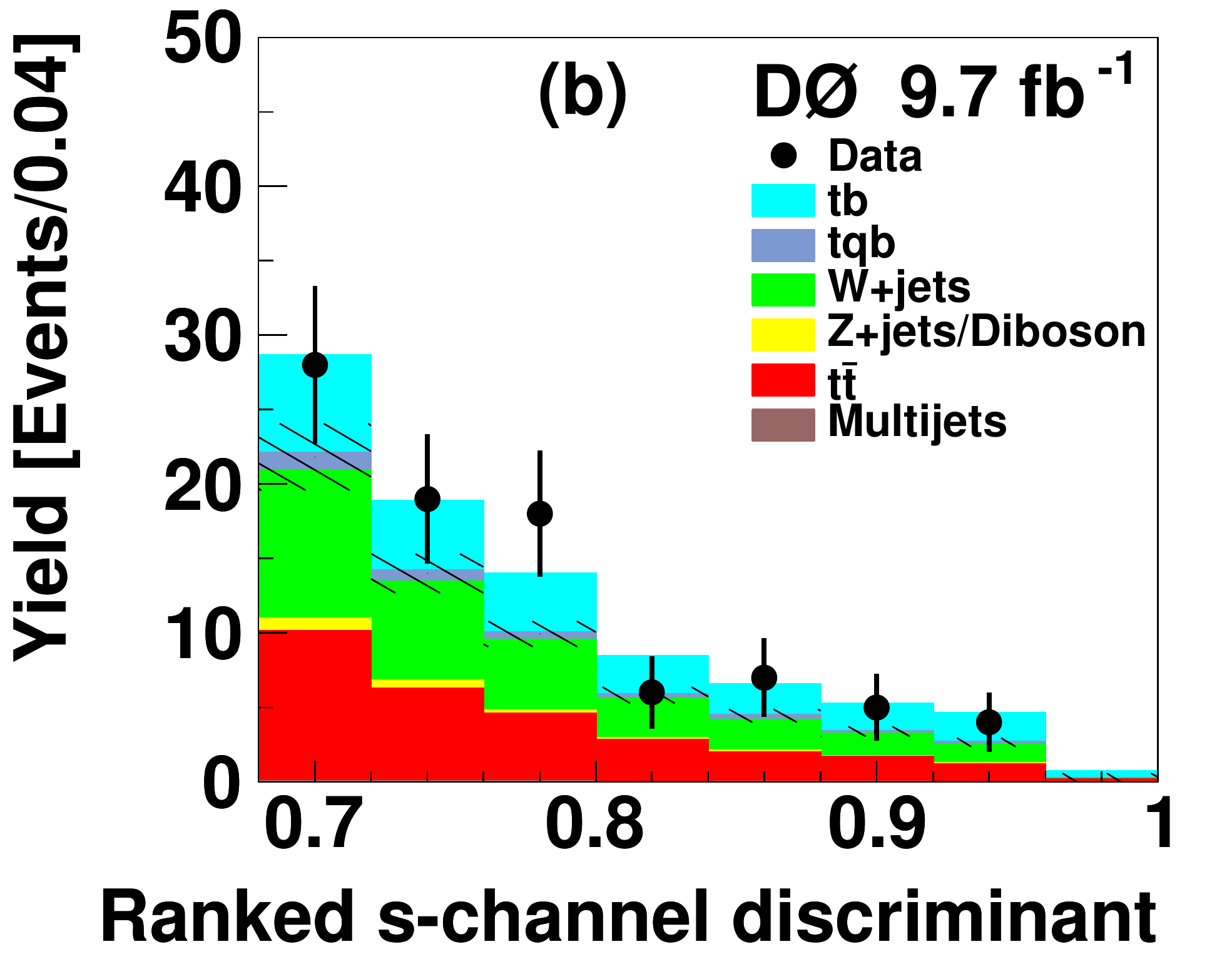}
\label{fig:T13AF3b}}
\caption{D0 lepton+jets discriminant in the t-channel (a) and s-channel (b) signal regions.}
\label{fig:figure}
\end{figure}

The D0 single top analysis is performed in the combined s+t sample using events selected in the lepton+jets final state. The tagging algorithm is based on a multivariate technique \cite{D0_btag}. Three multivariate methods are used: matrix element (ME), Bayesian Neural Network (BNN) and boosted decision tree (BDT). 
The correlation among the output of the individual MVA methods is measured to be around 75\%. 
A final BNN is built in order to construct combined discriminants for t- and s-channel signals ($D_t^{comb}$ and $D_s^{comb}$), shown in Fig.\ref{fig:T13AF3d} and Fig.\ref{fig:T13AF3b}, respectively.
The inclusive s+t single top cross section as well as the t- and s-channel individual cross section are extracted using a Bayesian approach. The resulting 2D posterior density distribution is shown in Fig.\ref{fig:T13AF4a}, together with several modes of new physics \cite{new_physics}. The measured value for the t-channel cross section is $\sigma_t = 3.07^{+0.54}_{-0.49}$. With no assumptions on the relative t- and s-channel contributions, the total single top cross section is measured to be $\sigma_{s+t} = 4.11^{+0.59}_{-0.55}$ pb\cite{D0_singletop}.

\subsection{Tevatron s+t Combination}
\begin{figure}[ht]
\centering
\subfigure[]{
\includegraphics[width=12pc]{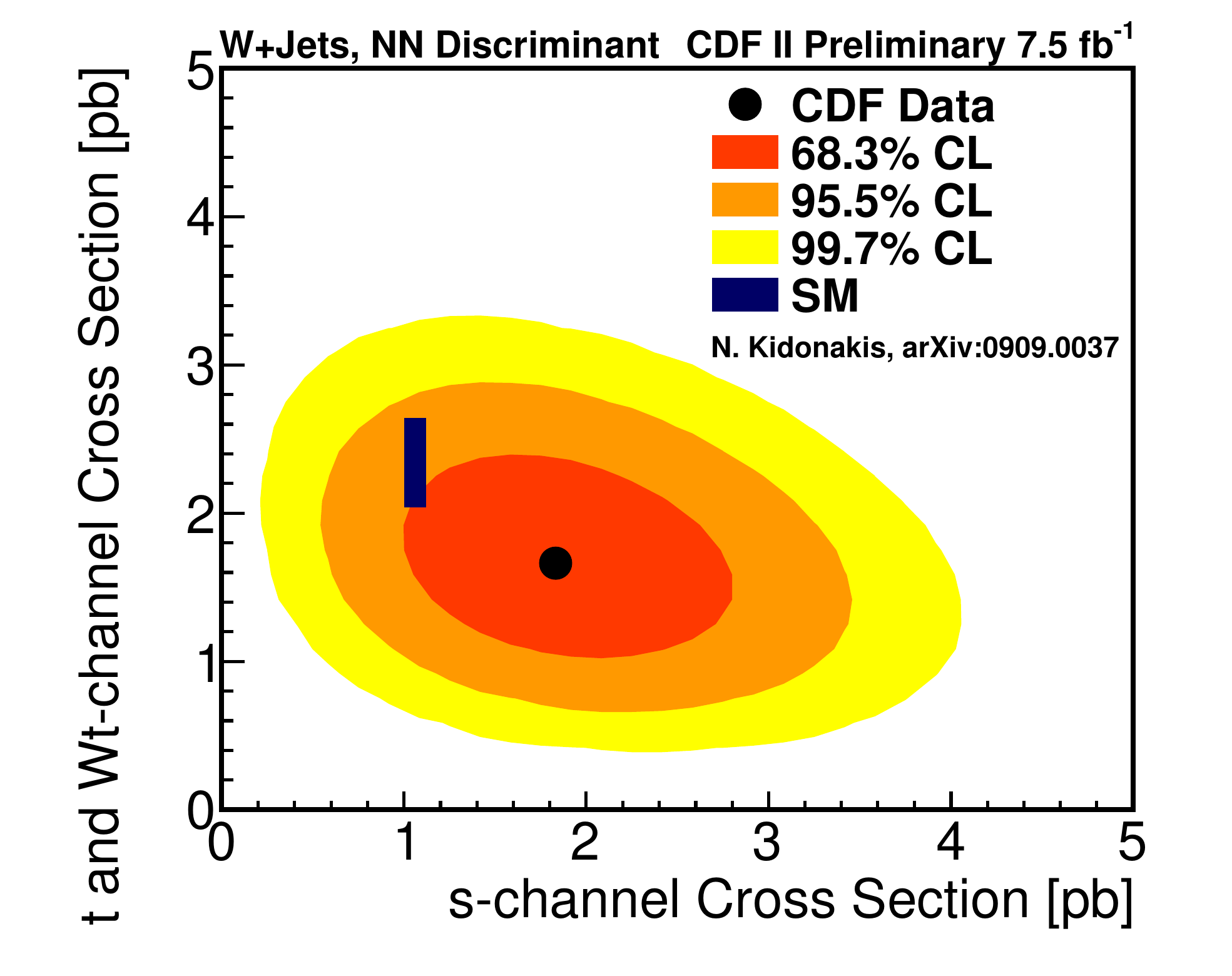}
\label{fig:2D_sVst}}
\subfigure[]{
\includegraphics[width=10pc]{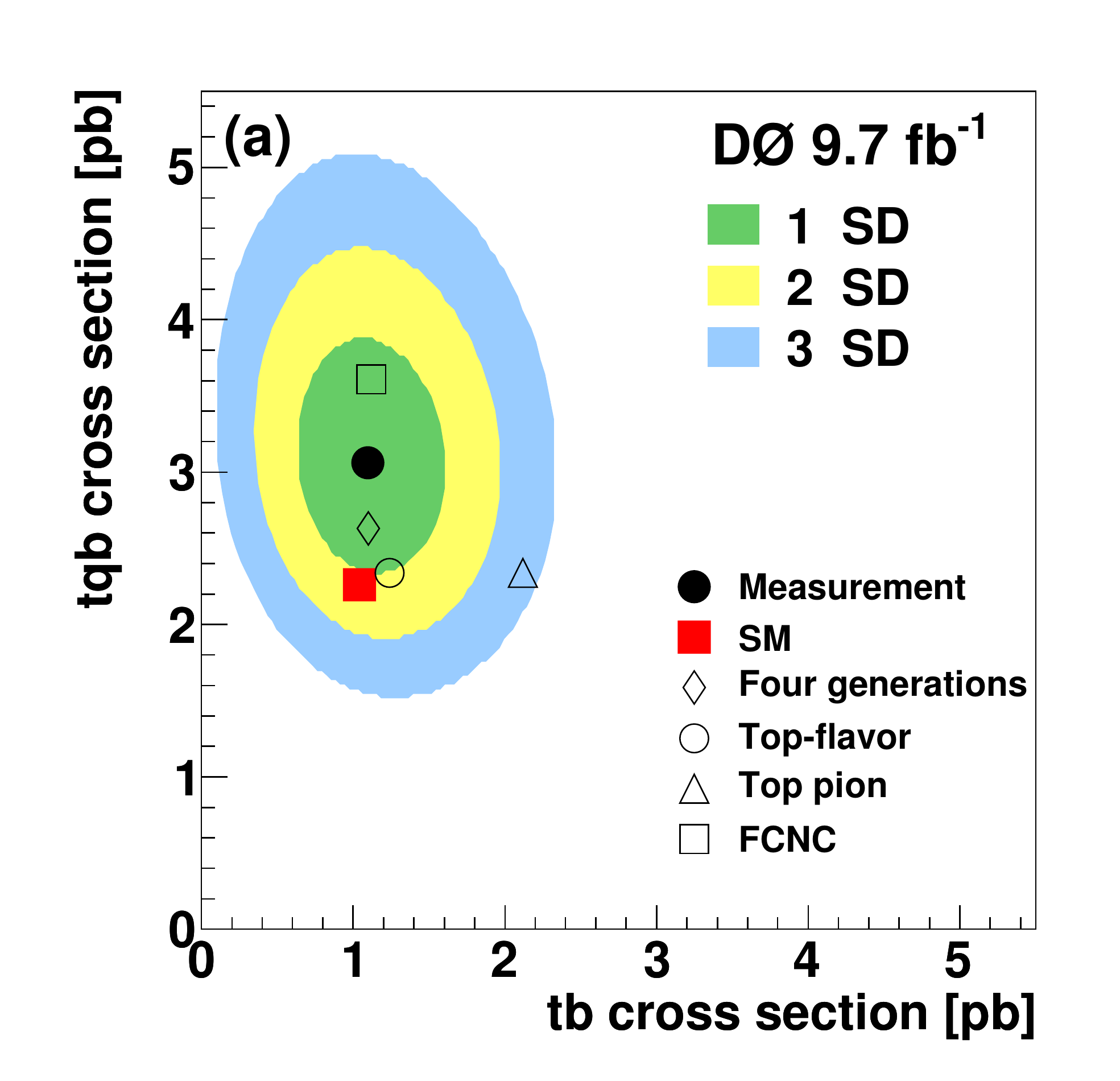}
\label{fig:T13AF4a}}
\subfigure[]{
\includegraphics[width=9pc]{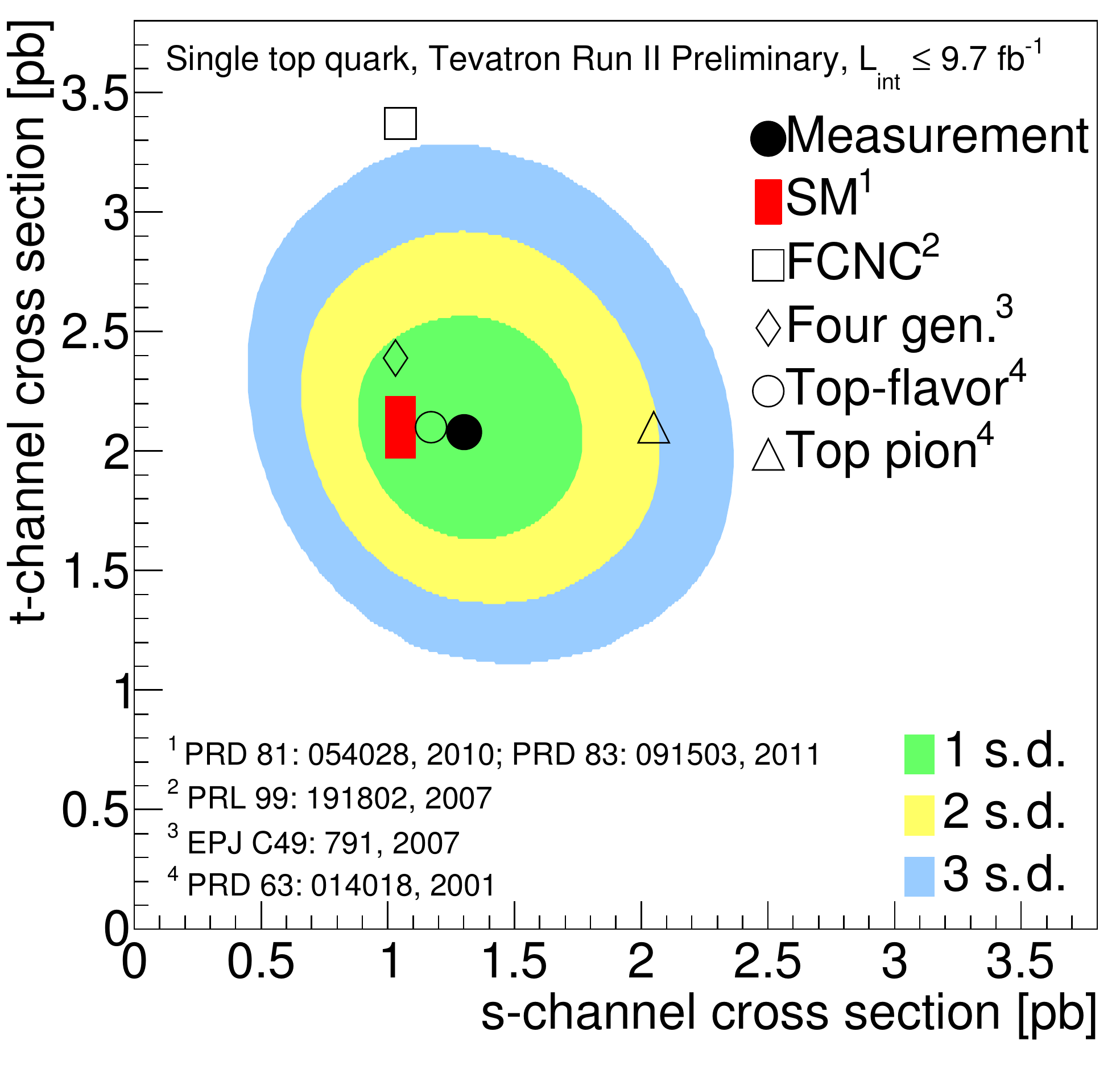}
\label{fig:Tev_2d_bsm}}
\caption{Posterior density as a function of t-channel and s-channel single top cross section for CDF l+jets analysis (a), for D0 analysis (b) and the Tevatron single top combination (c).}
\label{fig:figure}
\end{figure}

A combination of the CDF and D0 measurements
has been recently performed using a total integrated luminosity of up to 9.7 fb$^{-1}$. The combination utilizes the s- and t-channel discriminants from CDF and D0 single top quark measurements and forms a binned likelihood as a product of all analysis channels in the bins of the multivariate discriminants. The most probable value for the combined t-channel cross section is ￼$\sigma_t = 2.25^{+0.29}_{-0.31}$ pb. The combined s+t cross section is measured without assuming the SM ratio of $\sigma_s/\sigma_t$ to be
$\sigma_{s+t} = 3.30^{+0.52}_{-0.40}$ pb. The extracted 2D posterior probability distribution as a function of $\sigma_t$ and $\sigma_s$ is in Fig.\ref{fig:Tev_2d_bsm}\cite{Tev_st_Comb}.

\subsection{Extraction of the CKM matrix element $|V_{tb}|$}
The extraction of the CKM matrix element $|V_{tb}|$ from the single top cross section measurement proceeds via the formula $|V_{tb}|^2 = |V_{tb}^{SM}|^2 \times \frac{\sigma^{obs}}{\sigma^{SM}}$. No assumptions on the number of quark generation or the unitarity of the CKM is done. The resulting 95\% CL lower limit on $|V_{tb}|$ from the CDF combined measurement is 0.84 \cite{st_CDF_metjet} and the lower limit from D0 measurement is 0.92. The extracted limit from the Tevatron combined analysis is 0.92 at 95\% CL \cite{Tev_st_Comb}.
 
\section{Single top s-channel observation at Tevatron}
\begin{figure}[t]
\centering
\subfigure[]{
\includegraphics[width=13pc]{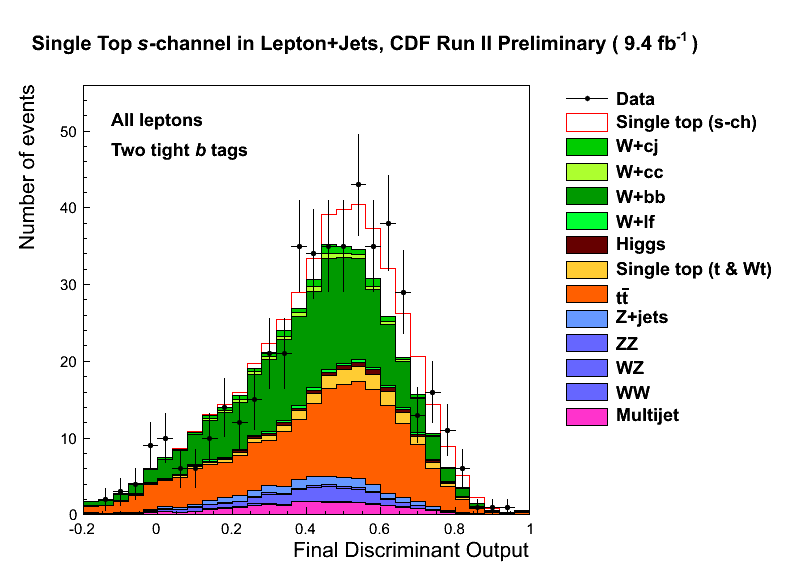}
\label{fig:ST_2JET_HobitTHobitT_CEMCMUPCMXISOTRK_bnnST}}
\subfigure[]{
\includegraphics[width=10pc]{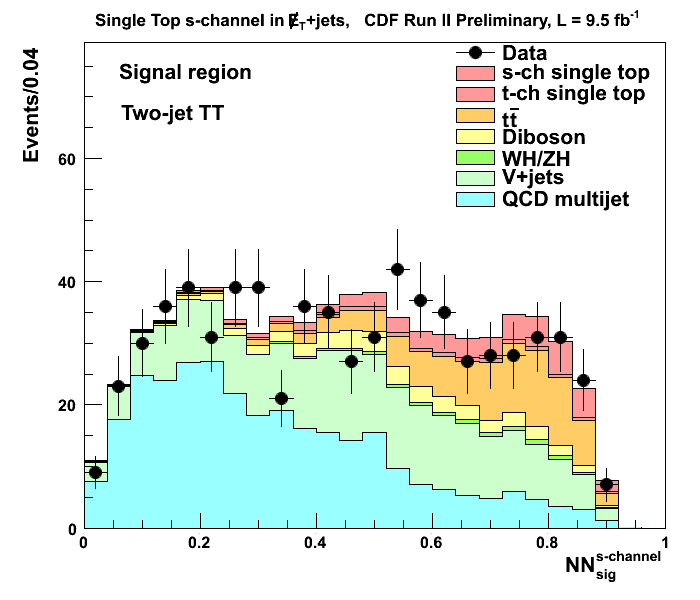}
\label{fig:Nvtx_sig_relax_schannel_2j_055_SIGNN2jTT_schannel_TT_CENT.png}}
\subfigure[]{
\includegraphics[width=9pc]{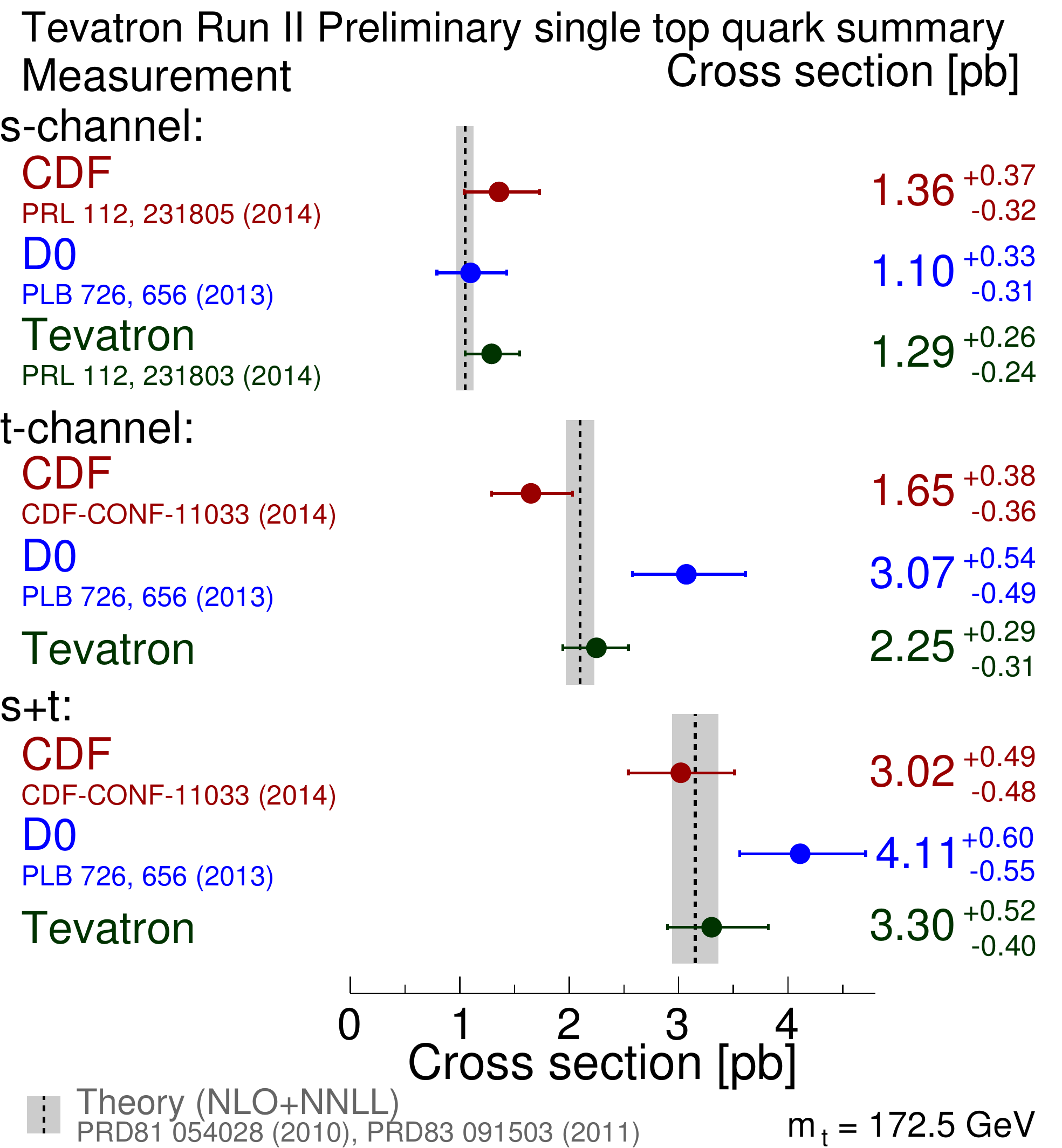}
\label{fig:tev_all_sum_plot}}
\caption{s-channel final discriminant in the CDF lepton+jets analysis (a) and \met+jets analysis (b). Summary of the Tevatron single top quarks cross section measurements (c).}
\label{fig:figure}
\end{figure}

At CDF, analyses optimized for the s-channel search have been performed in both lepton+jets and \met+jets final states, using the full CDF dataset (up to 9.5 fb$^{-1}$). 
The lepton+jets analysis forms an NN-discriminant to separate the signal. The same leptonic selection as in the inclusive single top quark search is used in the 2J2T signal region (with the same tagging algorithm as in \cite{st_CDF_metjet}). 
The discriminant distribution in the 2JTT signal region is shown in Fig.\ref{fig:ST_2JET_HobitTHobitT_CEMCMUPCMXISOTRK_bnnST} \cite{CDF_s-channel_lepjet}. 
In the \met+jets analysis the events are classified in the 2-3 jets region with different tagging category (T,TT,TL). The NN output for the \met+jets analysis in 2JTT category is shown in Fig.\ref{fig:Nvtx_sig_relax_schannel_2j_055_SIGNN2jTT_schannel_TT_CENT.png} \cite{CDF_s-channel_metjet_and_combined}.
The two CDF analyses are combined by taking the products of the likelihood and simultaneously varying the correlated uncertainties. A combined s-channel cross section of $\sigma_s = 1.36^{+0.37}_{-0.32}$ pb is measured, corresponding to a signal significance of 4.2 stardard deviations \cite{CDF_s-channel_metjet_and_combined}. 
At D0, the 2D posterior distribution shown in Fig.\ref{fig:T13AF4a} is also used to extract the cross section measurement for the s-channel. The measured cross section is $\sigma_s = 1.10^{+0.33}_{-0.31}$ pb, corresponding to a significance of 3.7 standard deviations \cite{D0_singletop}.

The Tevatron s-channel combination is performed using as inputs the CDF letpon+jets and \met+jets final discriminants and the D0 discriminant. The combination measures a cross section for the s-channel production of $\sigma_s = 1.29^{+0.26}_{-0.24}$ pb \cite{Tev_s-channel_and_combined}. The measurement corresponds to a significance of 6.3 standard deviations, reporting the first observation of the single top s-channel production (which is also the first observation made through a Tevatron combination).

\section{Conclusions}
We presented the most recent measurements of the single top quark production and the $|V_{tb}|$ matrix element at the Tevatron, which are summarized in Fig.\ref{fig:tev_all_sum_plot}. The combination between the CDF analyses and the previous D0 analysis on the single top s-channel cross section led to the first observation with an uncertainty of 19\% and a significance of 6.3 standard deviations. 
All the single top quark measurements at the Tevatron are in agreement with the SM predictions. With the recent finalization of the Tevatron combined s+t production measurement, the Tevatron single top program is mostly complete.


\end{document}